\documentclass[rnote]{aa}

\usepackage{graphicx}

\begin{document}

\title{H$\alpha$ spectropolarimetry  of RY~Tau and PX Vul\thanks{Based on observations obtained at the {\it Observat\'orio do Pico dos Dias}, LNA/MCT, Itajub\' a, Brazil.}}

\author{A. Pereyra\inst{1} \and A. M. Magalh\~aes\inst{2} \and  F. X. de Ara\' ujo\inst{1}}

\offprints{A. Pereyra, \email{pereyra@on.br}}

\institute{
Observat\'orio Nacional, Rua General Jos\'e Cristino 77, S\~ao Cristov\~ao, 20921-400, Rio de Janeiro, Brazil \and
Departamento de Astronomia, IAG, Universidade de S\~ao Paulo, Rua do Mat\~ao 1226, S\~ao Paulo, SP, 05508-900, Brazil}

\date{Received dd-mm-yy / Accepted dd-mm-yy}

\abstract
{}
{To detect line effects using spectropolarimetry in order to find evidence of rotating disks and their respective symmetry axes in T Tauri stars.}
{We used the IAGPOL imaging polarimeter along with the Eucalyptus-IFU to obtain spectropolarimetric measurements of the T Tauri stars \object{RY~Tau} (two epochs) and \object{PX~Vul} (one epoch). 
Evidence of line effects showing a loop on the {\it Q}$-${\it U} diagram favors a compact rather than an extended source for the line photons in a rotating disk. In addition, the polarization position angle (PA) obtained using the line effect can constrain the symmetry axis of the disk.}
{\object{RY~Tau} shown a variable H$\alpha$ double peak on 2004-2005 data. 
Polarization line effect is evident on the {\it Q}$-${\it U} diagram for both epochs confirming a clockwise rotating disk. A single loop is evident on 2004 changing to a linear excursion plus a loop on 2005. Interestingly, the intrinsic PA calculated using the line effect is consistent between our two epochs ($\sim$167$\degr$). An alternative intrinsic PA computed from the interstellar polarization corrected continuum and averaged between 2001-2005 yielded a PA$\sim$137$\degr$. This last value is closer to be perpendicular to the observed disk direction ($\sim$25$\degr$) as expected by single scattering in an optically thin disk. For \object{PX~Vul}, we detected spectral variability in H$\alpha$ along with non-variable continuum polarization when compared with previous data. The {\it Q}$-${\it U} diagram shows a well-defined loop in H$\alpha$ associated to a counter-clockwise rotating disk. The symmetry axis inferred by the line effect has a PA$\sim$91$\degr$ (with an ambiguity of 90$\degr$). Our results confirm previous evidence that the emission line in T Tauri stars has its origin in a compact source scattered off a rotating accretion disk.}
{}

\titlerunning{H$\alpha$ spectropolarimetry of RY~Tau and PX~Vul}
\authorrunning{Pereyra et al.}

\maketitle

\section{Introduction\label{intro}}

On the last years spectropolarimetry has become a powerful and unique tool for studying envelopes (or disks) around unresolved young stellar objects (YSOs). In this sense, the analysis of the Stokes parameters {\it Q}$-${\it U} diagram has shown to be a fundamental diagnostic tool for polarization line detections (Oudmaijer \& Drew~\cite{ou99}, Vink et al.~\cite{vi02,vi03} $-$ hereinafter V03,~\cite{vi05a}). 


Considering the line effects, the most simple case happens when the emission lines are formed over a much larger volume  than continuum. In this case, the circumstellar scattering material will polarize the continuum light more than the line photons. The last ones will add mainly unpolarized flux because they will be less scattered, and therefore the net polarization across the emission line will be reduced. This feature is usually named the depolarization line effect (Clarke \& McLean~\cite{cla74}) and a linear excursion is expected on the {\it Q}$-${\it U} diagram with the position angle (PA) unchanged. 


Models to explain the line effect including rotation also have been explored (Poeckert \& Marlborough~\cite{poe77}, Wood et al.~\cite{wo93}, Vink et al.~\cite{vi05b}). If the line photons originate from a compact source, the polarization pattern will depend on the specific geometry and bulk motions of the scattering particles surrounding the compact line source. In a rotating disk-like configuration, the subsequent breaking of left-right reflection symmetry in the velocity fields leads to a changing in polarization angle with wavelength, which appears as a loop on the 
{\it Q}$-${\it U} diagram (Vink et al. \cite{vi02}). Interestingly, the sense of rotation of the envelope/disk as seen from the Earth also can be retrieved of the direction in which the PA rotates around the loop on the {\it Q}$-${\it U} diagram (Poeckert \& Marlborough~\cite{poe77}). 


The survey of Vink et al.~(\cite{vi05a}) found that the most (9/10) of the T Tauri stars (TTS) shown line effect associated with a compact source of line photons that is scattered off a rotating accretion disk. This means that loops on the {\it Q}$-${\it U} diagram must be a well-defined signature for TTS. The statistics clearly has to be increased to confirm this issue. 


In this letter, we report spectropolarimetric observations around H$\alpha$ of the TTS \object{RY~Tau} and \object{PX~Vul}. These are the first results of an ongoing spectropolarimetric survey of YSOs. The observations and data reduction are presented in \S~\ref{data}. The results are shown in \S~\ref{resul}. A summary with the final conclusions are drawn in \S~\ref{concl}.

\section{Observations\label{data}}

\begin{figure}
\resizebox{8.8cm}{!}{
$\begin{array}{ll}
\resizebox{8.8cm}{!}{\includegraphics[clip]{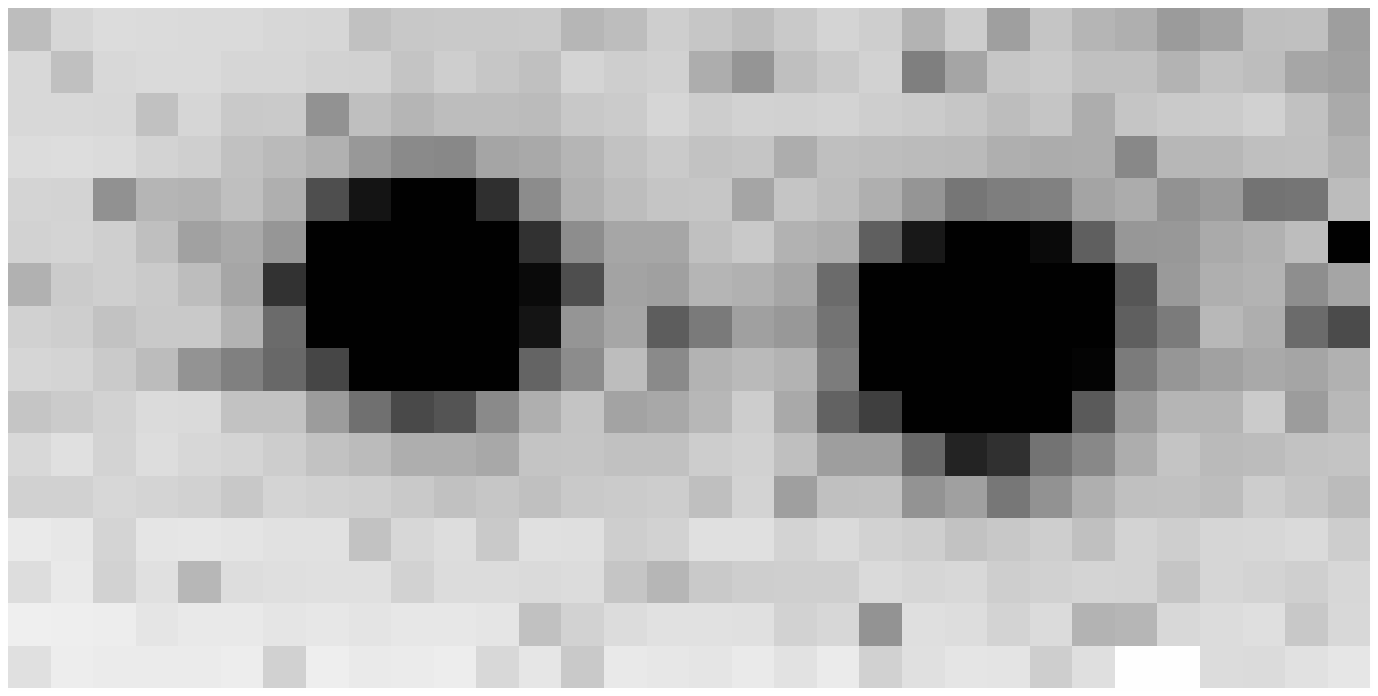}} &
\resizebox{8.8cm}{!}{\includegraphics[clip]{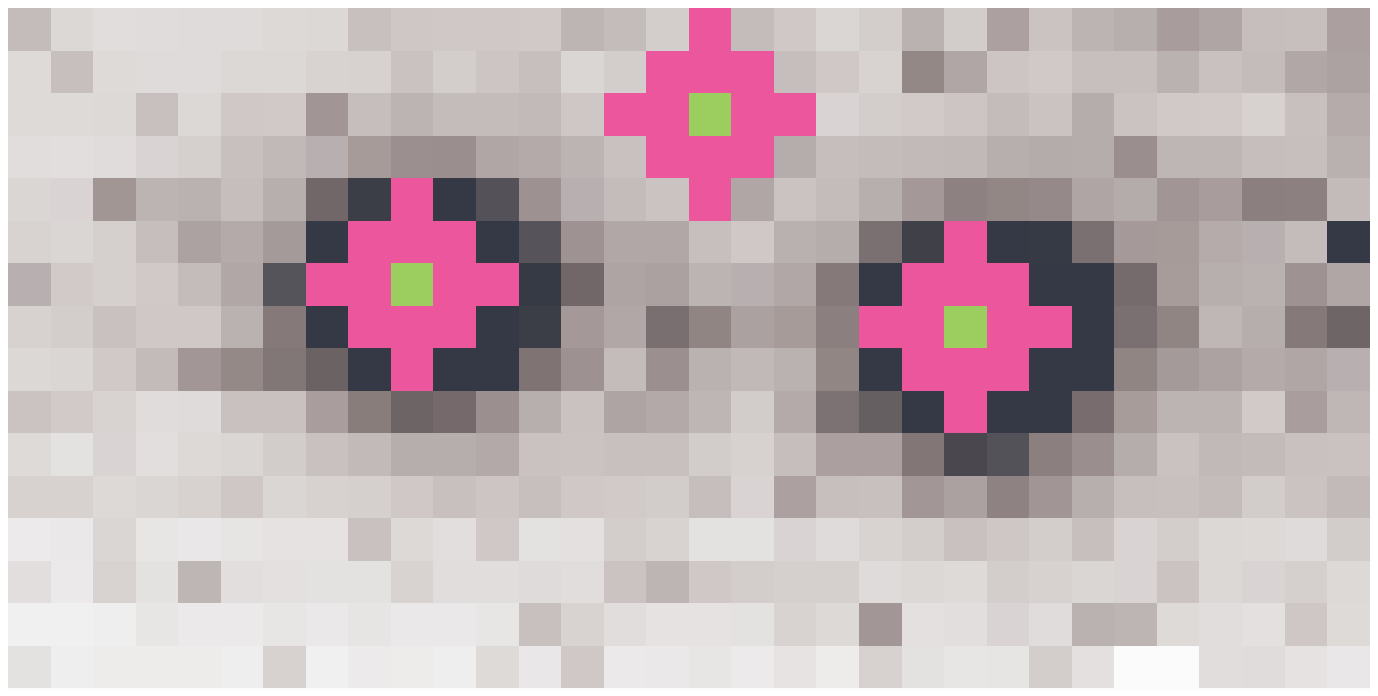}} \\
\end{array}$
\put (-441,-20) {\parbox[l]{0cm}{\Large EB}}
\put (-338,-29) {\parbox[l]{0cm}{\Large OB}}
\put (-311,55) {\parbox[l]{0cm}{\Large 30$\arcsec\times$15$\arcsec$}}
\put (-186,-20) {\parbox[l]{0cm}{\Large EB}}
\put (-84,-29) {\parbox[l]{0cm}{\Large OB}}
\put (-108,55) {\parbox[l]{0cm}{\Large sky}}
\put (-245,-55) {\parbox[l]{8.8cm}{\large Extracted fibers (aperture radius: 2pix = 2$\arcsec$)}}
}
\caption{\it{Left}\rm: 2-D reconstructed image from EIFU+IAGPOL showing the ordinary (OB) and extraordinary (EB) beams for point sources in one of the waveplate positions. The field size is 30$\arcsec\times$15$\arcsec$ with an scale of 0.93$\arcsec$ per pixel. 
\it{Right}\rm: Extracted fibers (in magenta) used to construct the ord., ext. and sky spectra in each waveplate position for an appropriate aperture radius. In this example, $\it{r}$=2pix or 13 stacked fibers. The central fiber is shown in green for each spectra.}
\label{eifu}
\end{figure}

The observations were made in two runs on 2004 and 2005 using the 1.6m telescope at the Observat\'orio do Pico dos Dias (OPD), Brazil. We used IAGPOL, the IAG imaging polarimeter (Magalh\~aes et al. \cite{ma96}, Pereyra ~\cite{pe00}), along with the Eucalyptus-IFU (EIFU, de Oliveira et al. \cite{deol03}). EIFU is an integral field unit composed of an array of 32~$\times$~16~50~$\mu$m fibers that covers a field of 30$\arcsec~\times$~15$\arcsec$ on sky with a scale of 0.93$\arcsec$ per pixel (Fig.~\ref{eifu}, left). The detector used was a Marconi 2048~$\times$~4608 pixels back-illuminated CCD with 13.5~$\mu$m$^{2}$ per pixel. The used spectral range for EIFU was $\sim$600~$\rm \AA$ centered in H$\alpha$ ({\it R}=4000) that yields a resolution of $\sim$0.3~$\rm \AA$ per pixel. 

We used IAGPOL in linear polarization mode with a Savart plate as analyzer and an achromatic $\lambda$/2-waveplate as a retarder. Each measurement consisted in four or eight waveplate positions separated by 22$\fdg$5. The IAGPOL$+$EIFU setup permits that the beam collected by the telescope and divided in two orthogonal polarization beams by IAGPOL will be projected on the EIFU fibers array. Then, spectropolarimetry can be done using the ordinary and extraordinary beams that yield the -o and -e spectra in each waveplate position, respectively. A log of observations is shown in Table~\ref{log}. 

We used standard $\it{IRAF}$\footnote{$\it{IRAF}$ is distributed by the National Optical Astronomy Observatory, which is operated by the Association of Universities for Research in Astronomy, Inc., under cooperative agreement with the National Science Foundation.} procedures for IFU reductions in each waveplate position image including bias and flatfielding corrections along with wavelength calibrations. Then, a special routine was developed to extract and stack fibers for the -o and -e beams and the sky region (Fig.~\ref{eifu}, right). The optimum aperture radius for the fibers extraction was selected by minimizing the polarimetric errors. Therefore, three spectra were constructed in each waveplate position: O($\lambda$), E($\lambda$) and sky($\lambda$) where special care was taken to remove cosmic rays. The sky spectrum was constructed stacking fibers away from the -o and -e beams (Fig.~\ref{eifu}, right) and subtracted of the -o and -e spectra in each waveplate position image. Several relative positions for the sky were tested with similar results. After that, we used SPECPOL$\footnote{the original version of SPECPOL was written by A. Carciofi.}$ package to construct the flux, polarization, polarization PA and polarized flux spectra for a proper binning. This package allows binning the spectra using a variable bin size set by a selected and constant polarization error per bin. The errors are obtained from the residuals of the observations at each waveplate position image with respect to the expected cosinusoid curve. In general, they are consistent with the photon statistics.

Observations of polarized standard stars in each run yielded the position angle correction to the equatorial system ($\rm PA_{corr}$) and permitted to check the polarization efficiency of our setup. Our measurements are in agreement with the literature values within one sigma. In addition, the instrumental polarization was checked using unpolarized stars and estimated to be lower than 0.2$\%$, and therefore no correction was needed. A summary of the calibration data is shown in Table~\ref{cal}. Typical signal-to-noise ratios (per~bin) gathered by our setup for the sources in this work are between 170 and 250. We checked for the polarization bias effect and it shown to be negligible.

\begin{table}
\caption{Log of observations.}
\begin{tabular}{lllll}
\hline \hline
Object		&   $\it{V}^{\rm a}$ 	&  Date 			& N$^{\rm b}$  	& Total IT$^{\rm c}$\\
          &    (\rm mag)        &             &             & (s) \\       
\hline
\multicolumn{5}{c}{T Tauri}\\
\hline
RY Tau  	&		10.2 	& 2004/sep/05 & 8 	& 2400 \\
					&		      & 2005/sep/03 & 4 	& 2400 \\
PX Vul		&		11.7	  & 2005/sep/03 & 8 	& 4800 \\
\hline
\multicolumn{5}{c}{standards}\\ 
\hline
HD 147084$^{\rm d}$ &		4.6 	& 2004/sep/05 & 8 	& 1600 \\
HD 187929$^{\rm d}$	&		3.9		& 2005/sep/03 & 8 	& 640  \\
HD 10476$^{\rm e}$	&		5.2		& 2005/sep/03 & 8 	& 800  \\
\hline
\end{tabular}
\\
(a) from Simbad; (b) number of waveplate positions; (c) total integration time (N $\times$ individual IT by waveplate position); (d) polarized; (e) unpolarized.\\
\label{log}
\end{table}

\begin{table}
\scriptsize
\caption{Calibration summary.}
\begin{tabular}{llrlll}
\hline \hline
Object		& $\it{P}_{\rm obs}$$^{\rm a}$ & PA$_{\rm obs}$$^{\rm a}$ & $\it{P}_{\rm lit}$$^{\rm b}$ & PA$_{\rm lit}$$^{\rm b}$ & PA$_{\rm corr}$\\
          &     (\%)      &   (\degr)      &    (\%)       &   (\degr)      & (\degr)    \\ 
\hline
HD 147084 &		4.5 (0.2) & 0.2          & 4.50 (0.01)   &   33.2         & $+$33.0  \\
HD 187929	&		1.5 (0.2) & 37.9         & 1.62 (0.05)   &   91.0         & $+$53.1  \\
HD 10476	&		0.2 (0.2) & --           & 0.02 (0.01)   &      --        &   --     \\
\hline
\end{tabular}
\\
errors in parenthesis.
(a) mean values for a spectrum binned with a polarization error (per bin) of 0.2\%, excepts HD10476 (unbinned). The quoted polarization error is the averaged error per bin.
(b) References: HD 147084, Tapia~(\cite{tap88}), $\it{R}$ filter; HD 187929, Bailey \& Hough~(\cite{bai82}), $\it{R}$ filter; HD 10476, Bastien et al. (1988), $\it{V}$ filter. 
\label{cal}
\end{table}

\section{Results}
\label{resul}

\subsection{RY Tau}
\label{rytau}

\object{RY~Tau} is a classical TTS (F8III type, Mora et al. \cite{mo01}) 
with a high optical variability 
and possibly associated with variable circumstellar obscuration as in UX~Ori type (Beck \& Simon \cite{be01}). 
Spectral variability also has been observed (Vrba et al. \cite{vr93}, Petrov et al. \cite{pe99}) without a clear  correlation between the H$\alpha$ line strength and the optical magnitudes. 

\object{RY~Tau} shows a historical strong linear polarization variability (Hough et al. \cite{ho81}, Bastien \cite{bas82}, Bergner et al. \cite{ber87}, V03 and references therein). 
The proposed polarizing mechanisms include rotational modulations by cool and hot spots on the stellar surface (Stassun \& Wood \cite{sta99}) and variable extinction by a dusty disk (M\'enard \& Bastien \cite{me92}). Recently, robust evidence of a bipolar jet (PA$_{\rm jet}$=115$\degr$) was found in \object{RY~Tau} (St-Onge \& Bastien~\cite{sto08}) using H$\alpha-$continuum imaging that can be used to infer the direction of the unresolved disk and constrain the polarimetric information.

\begin{table}
\caption{Results.}
\begin{tabular}{llllr}
\hline \hline
Object		& Date				& $\it{P}_{\rm obs}$$^{\rm a}$	& PA$_{\rm obs}$$^{\rm a}$	&   PA$_{\rm int}$	\\
          &							&  (\%)								&   (\degr)				&    (\degr)    		\\ 
\hline
RY Tau  	&	2004/sep/05	&    1.8 (0.3) 				& 11 (5)       		&   171 (11)  \\
      		&	2005/sep/03	&    2.6 (0.2)				& 16 (2)       		&   161$^{\rm b}$ (4)		\\
PX Vul		&	2005/sep/03	&    3.2 (0.3)				& 33 (2)       		&   91$^{\rm c}$ (25)	\\
\hline
\end{tabular}
\\
errors in parenthesis. 
(a) mean values for a spectrum binned with a polarization error (per bin) of 0.3$\%$. The quoted polarization error is the averaged error per bin. 
(b) rebinned at 0.15\% per bin.
(c) or 91$+$90.
\label{tabresults}
\end{table}

\begin{figure*}
$\begin{array}{cc}
\resizebox{8.8cm}{!}{
\resizebox{8.8cm}{8cm}{\includegraphics{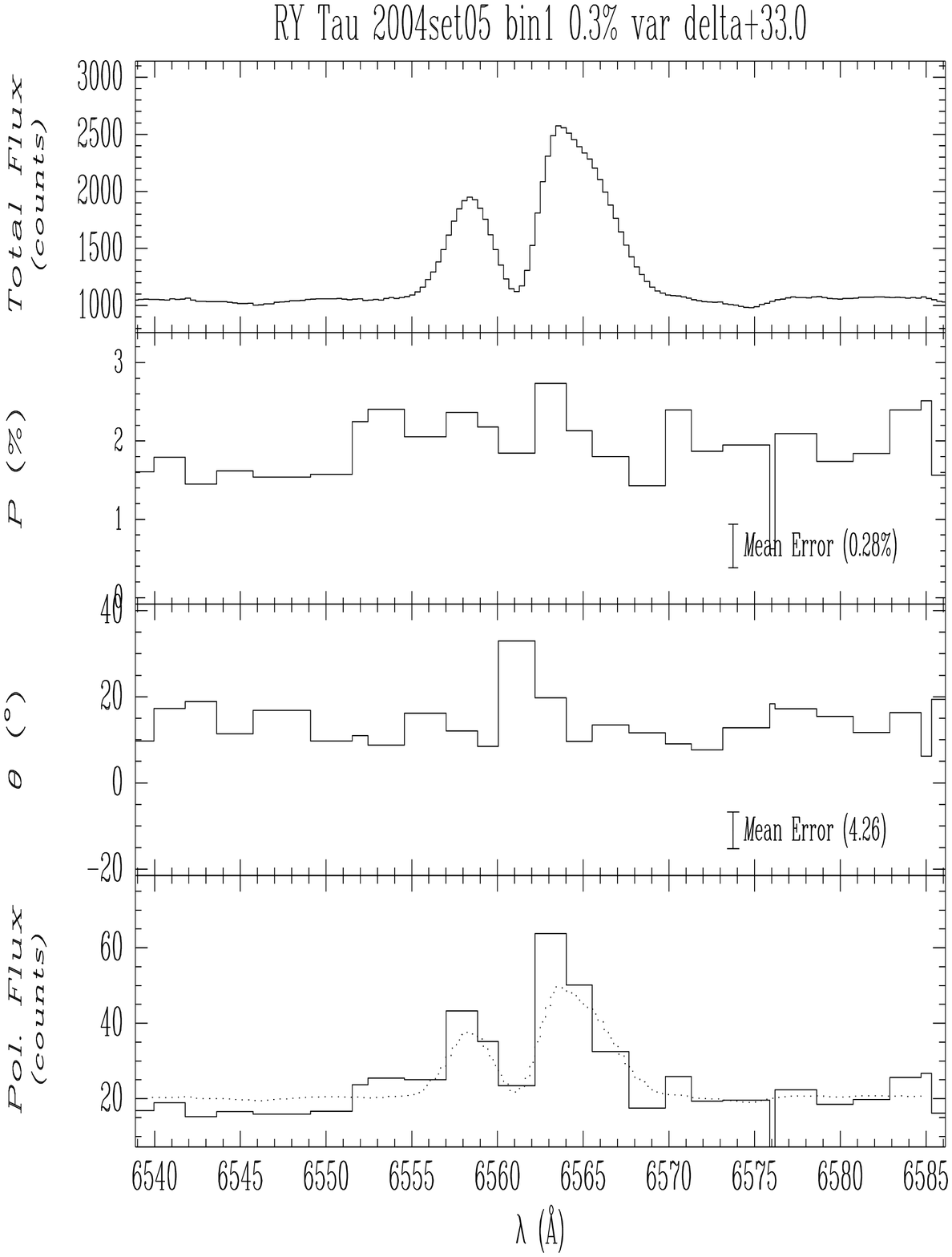}}
\linethickness{0.1pt} 
\put (-119.6,210) {\line(0,-1){185}}
\put (-97.0,210) {\line(0,-1){185}}
}
&
\resizebox{8.8cm}{!}{
\resizebox{8.8cm}{8cm}{\includegraphics{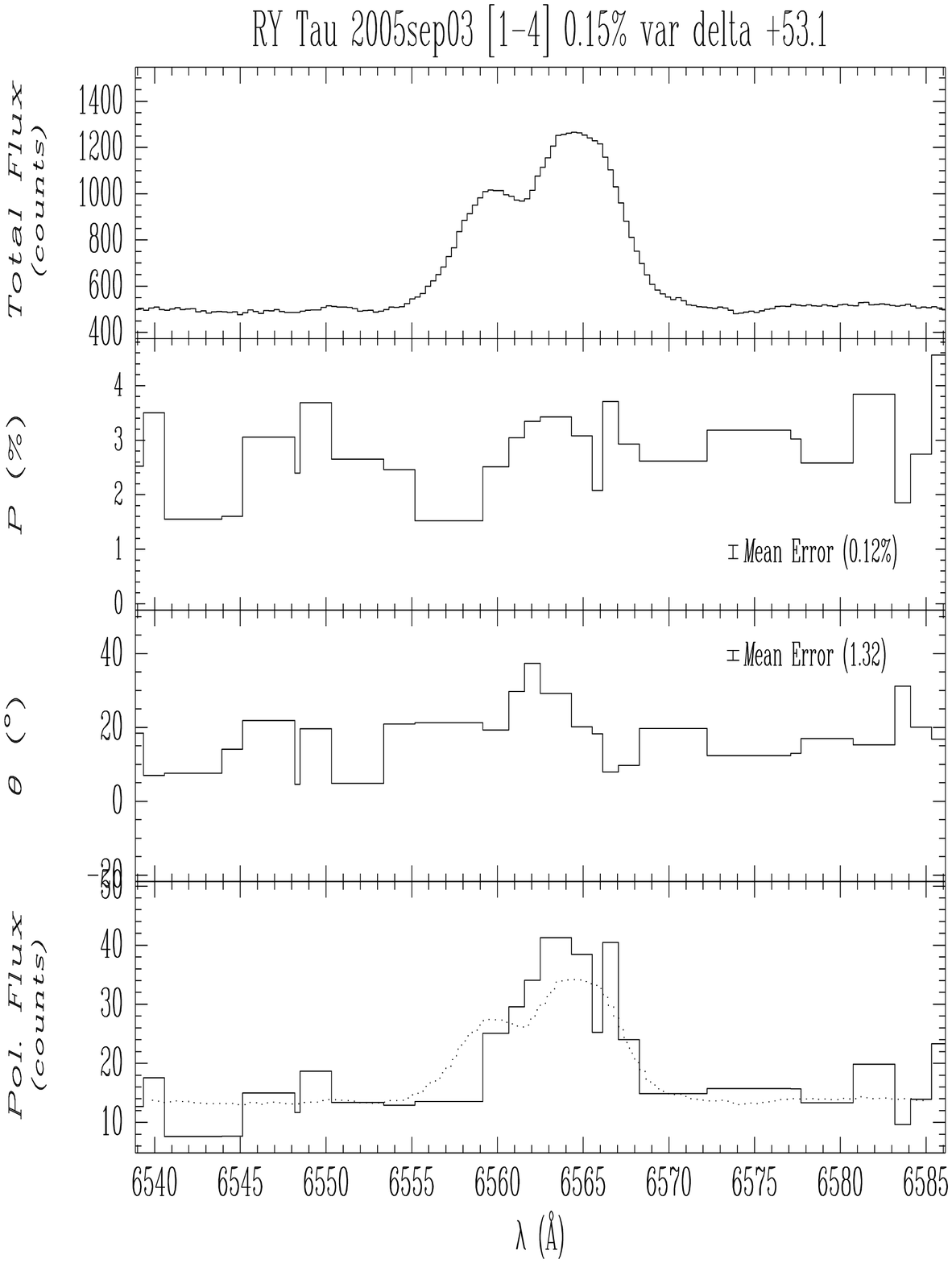}}
\linethickness{0.1pt} 
\put (-119.6,210) {\line(0,-1){185}}
\put (-97.0,210) {\line(0,-1){185}}
}
\\
\resizebox{4.4cm}{!}{\includegraphics{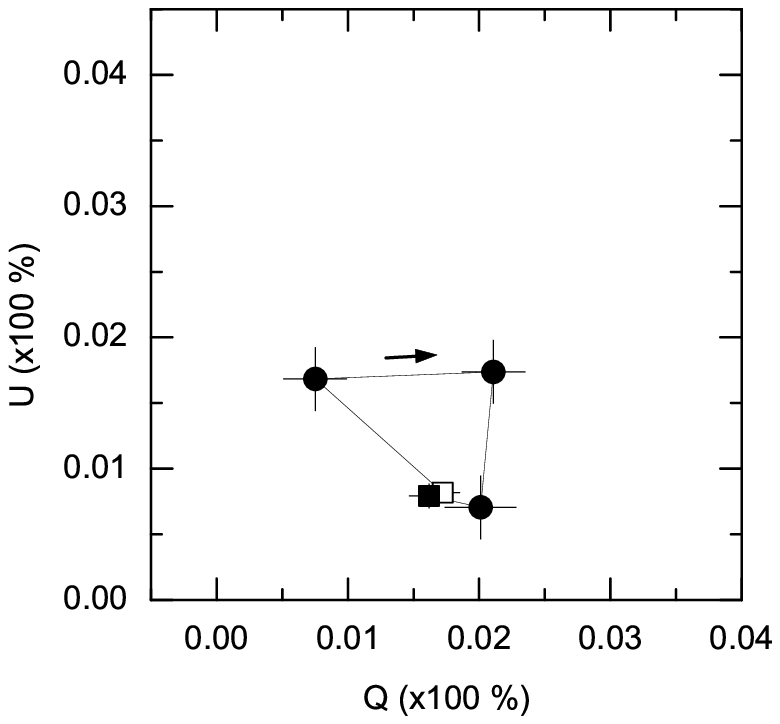}}
&
\resizebox{4.4cm}{!}{\includegraphics{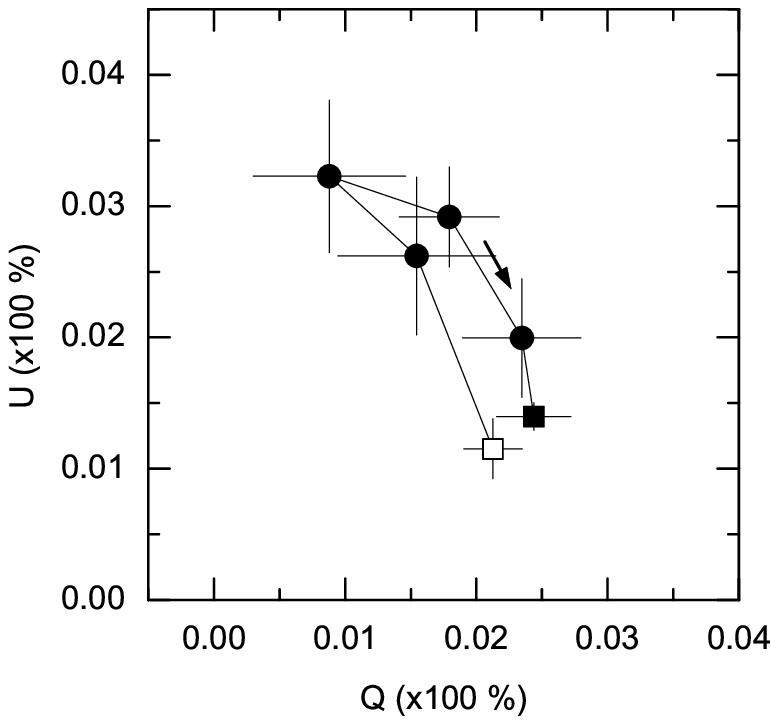}}
\\
\end{array}$

\caption{Spectropolarimetry of \object{RY~Tau} around H$\alpha$ on 2004/sep/05 (left) and 2005/sep/03 (right). \it{Top}\rm: The vertical lines indicate the region (6560--6565~$\rm \AA$) where the line effect is evident. The spectra show the total flux (top), polarization (upper middle), polarization PA (lower middle), and polarization flux (bottom) binned using a variable bin size with a constant polarization error (per bin) of 0.3$\%$ and 0.15$\%$ for the 2004 and 2005 data, respectively. \it{Bottom}\rm: $\it{Q}$$-$$\it{U}$ diagram showing the line effect (black dots). The direction in which the wavelength increases is also indicated (arrow) along with the mean values for the blue (white square) and red (black square) continuum. \textbf{The errors plotted around the line effect are assuming photon statistics directly.}}
\label{linerytau1}
\end{figure*}


Table~\ref{tabresults} shows the observed continuum polarization for \object{RY~Tau} integrated along the full range of our spectra. A slight enhancement of the continuum polarization level is observed with the PA practically unchanged ($\sim$14$\degr$) in one year. Our values are comparable with the data from Vink~et~al.~(\cite{vi05a}) between 2001 and 2003. The spectropolarimetry around H$\alpha$ is indicated in Fig.~\ref{linerytau1}. The spectral range of the top spectra is only of 45$\rm \AA$ for a better visualization of the line feature. The spectra are binned using a variable bin size with a constant polarization error (per bin) of 0.3$\%$ and 0.15$\%$ for the 2004 and 2005$\footnote{The Stokes parameter solution with few waveplate positions (as in 2005 data) trends to underestimate the polarization error (per bin). In order to detect the line effect, the 2005 data was rebinned at a level that guaranteed the best compromise between the line effect detection and the minimum dispersion of the continuum polarization.}$ data, respectively. These errors are the best compromise between a minimum error (per bin) and line effect detection. The lower plots are the $\it{Q}$$-$$\it{U}$ diagrams showing the line effect.

On 2004, the total flux of H$\alpha$ appeared as a double-peaked profile with the central dip at the continuum level and the redder peak more intense. On the other hand, the 2005 data again shown a similar double-peak but with the central dip just below the blue peak. The AVVSO 
database (Henden~\cite{he07}) shown that \object{RY~Tau} in average faded from $\it{V}$ $\sim$10.0 mag on 2004/sep to $\sim$10.8 mag on 2005/sep. The relative intensities of the H$\alpha$ emission components remained practically constant between the two epochs and can be classified as II$-$B type (Reipurth, Pedrosa \& Lago \cite{rei96}). We did not find evidence of an increment of H$\alpha$ when the star is dimmed (Petrov et al.~\cite{pe99}).

We detected the H$\alpha$ polarization line effect for \object{RY~Tau} using the $\it{Q}$$-$$\it{U}$ diagrams (Fig.~\ref{linerytau1}, bottom) around the central dip (between 6560--6565~$\rm \AA$). This wavelength range is also indicated on each spectrum (Fig.~\ref{linerytau1}, top) for a better comparison. On 2004, the line effect produces a loop already reported by V03. Nevertheless, the 2005 data show a stronger and more complex pattern that resembles a mixing between a loop and a linear excursion. The integrated continuum polarization immediately before and after of the line effect is the same in each epoch. Interestingly, the direction in which the wavelength increases around the loop (indicated by the arrow) has the same sense in our two epochs (and also in V03). Therefore, this confirms that the disk of \object{RY~Tau} rotates in a clockwise direction as seen by an observer at the Earth.  

In order to determine the intrinsic PA (PA$_{\rm int}$) we used a linear fit on the points that display the line effect on the $\it{Q}$$-$$\it{U}$ diagram. Formally, the sloop of the fit yields two possible solutions (PA$_{\rm int}$ and PA$_{\rm int}$+90$\degr$). In general, this ambiguity is solved by the solution more consistent with the relative positions of the interstellar polarization (ISP) and the observed polarization on the $\it{Q}$$-$$\it{U}$ diagram (Schulte-Ladbeck et al.~\cite{sc94}). 

\begin{figure}
\resizebox{8.8cm}{!}{\includegraphics{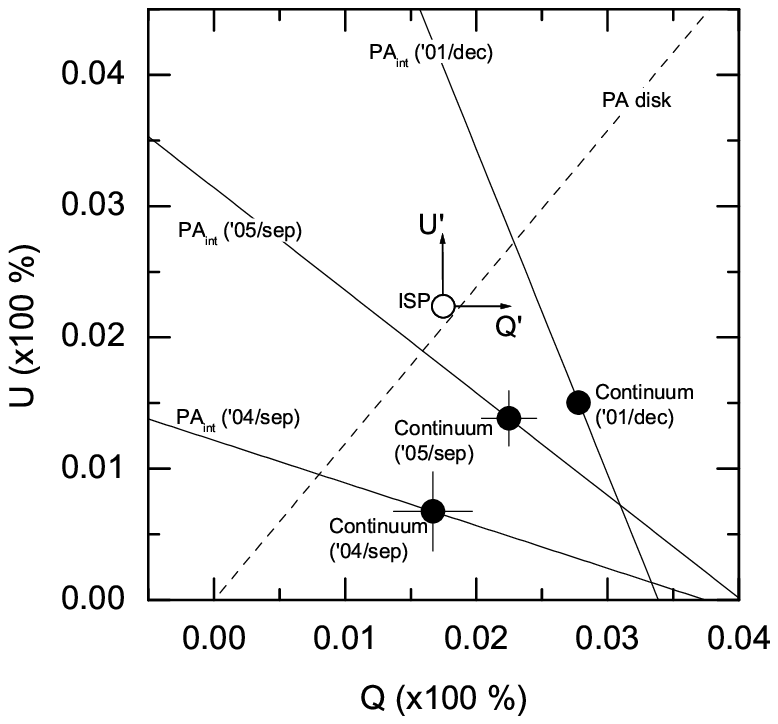}}
\caption{$\it{Q}$$-$$\it{U}$ diagram for RY Tau showing the intrinsic polarization PA (solid lines) from the H$\alpha$ line effect on three epochs (2001/dec from Vink et al.~\cite{vi03}; and, 2004/sep and 2005/sep from this work). The continuum polarization around H$\alpha$ also is shown in each case (black dots). The interstellar polarization (Petrov et al.~\cite{pe99}) is also indicated (white dot) along with the PA disk (dashed line, St-Onge \& Bastien~\cite{sto08}). The arrows indicate the reference system (Q',U') corrected by the ISP.} 
\label{rytaucompqu}
\end{figure}

Fig.~\ref{rytaucompqu} shows the observed continuum polarization and the PA$_{\rm int}$ directions obtained from the line effect for our two epochs (Table~\ref{tabresults}). The ISP toward \object{RY~Tau} (2.8$\%$ at PA=26$\degr$) computed by  Petrov et al.~(\cite{pe99}) is also shown. Analysing this diagram it seems clear that the solutions with $\it{U}^{'}$ $<$ 0 for PA$_{\rm int}$ are more likely. These solutions are indicated on the last column of Table~\ref{tabresults}. As we can see, the PA$_{\rm int}$ values are practically the same ($\sim$167$\degr$) between our two epochs considering the individual errors. As mentioned above, the same procedure applied by V03 on 2001 data yielded a PA$_{\rm int}\sim$146$\degr$ (also shown in Fig. ~\ref{rytaucompqu}) that is consistent with our results within $\sim$20$\degr$. Therefore, the three existing measurements using the line effect in H$\alpha$ spectropolarimetry yielded a mean PA$_{\rm int}\sim$160$\degr$ with a variation range of $\sim$15$\degr$.

Assuming the ISP from Petrov et al.~(\cite{pe99}) and discounting this value of the continuum polarization (Table~\ref{tabresults}), the intrinsic polarization is 1.6$\%$ at PA=134$\degr$ and 1.0$\%$ at PA=149$\degr$ on our 2004 and 2005 data, respectively. Vink et al.~(\cite{vi05a}) reported continuum polarization around H$\alpha$ in four epochs between 2001 and 2003. Therefore, considering the ISP used above, the mean intrinsic polarization for RY Tau between 2001 and 2005 is 1.3\% at PA=137$\degr$. The variation range of the PA$_{\rm int}$ is $\sim$25$\degr$ in this interval. 

Comparing the mean values for the PA$_{\rm int}$ by the above two methods, we obtained that they are reasonably consistent. In particular, the ISP corrected continuum method yields a PA$_{\rm int}$ closer to be perpendicular to the apparent disk direction (PA$_{\rm disk}$=25$\degr$) derived by the jet (St-Onge \& Bastien~\cite{sto08}). This fact reinforces the idea of an optically thin disk in \object{RY~Tau} if single scattering is the prevailing mechanism. It is interesting to note the apparent alignment between the disk direction and the ISP (see Fig.~\ref{rytaucompqu}) suggests a weak interplay between the local magnetic field and its presumed collimator effect on the jet (De Colle \& Raga~\cite{dec05}).

\subsection{PX Vul}
\label{pxvul}

\object{PX~Vul} (\object{LH$\alpha$~483-41)} was initially cataloged as an early type star with circumstellar shell and H$\alpha$ in emission at a distance of 420 pc and belonging to the R association \object{Vul~R2} (Herbst et al.~\cite{her82}). Its position is coincident with the edge of the filamentary cloud \object{L778} near its northern NH$_{3}$ core (Myers et al. \cite{mye88}). Mora et al. (\cite{mo01}) classified \object{PX~Vul} as an early TTS (F3 Ve) and Hernandez et al. (\cite{her04}) as a Herbig Ae/Be star (F3$\pm$1.5). 
Nevertheless, the last inferred masses ($\leq$2M$\sun$) by Eisner et al. (\cite{eis05}) and Manoj et al. (\cite{man06}) seem more consistent with a TTS. 

Optical photometric variability ($\Delta\it{V}$ $\sim$0.3 mag) was reported by Herbst et al. (\cite{her82}). 
Nevertheless, Eiroa et al. (\cite{eir02}) classified \object{PX~Vul} as non-variable ($\Delta\it{V}$ $<$0.05 mag). 
This object shows a moderate veiling level ($\it{r_{R}}$~=0.8$\pm$0.4), an important rotational velocity {$\it v$}sin{\it i}~=78$\pm$11~km~s$^{-1}$, and a higher mass-accretion rate of 1.3$\times$10$^{-6}$~M$\sun$ yr$^{-1}$ (Eisner et al. \cite{eis05}).


The optical polarization of \object{PX~Vul} was not found to be variable by Oudmaijer et al. (\cite{ou01}). Their mean $\it{R}$ broadband polarization (3.8$\%$ at PA=27$\degr$) is comparable to our observed polarization integrated along the full spectrum (Table~\ref{tabresults}), confirming therefore the polarimetric non-variability. 

\begin{figure}
$\begin{array}{cc}
\resizebox{8.8cm}{!}{
\resizebox{8.8cm}{8cm}{\includegraphics{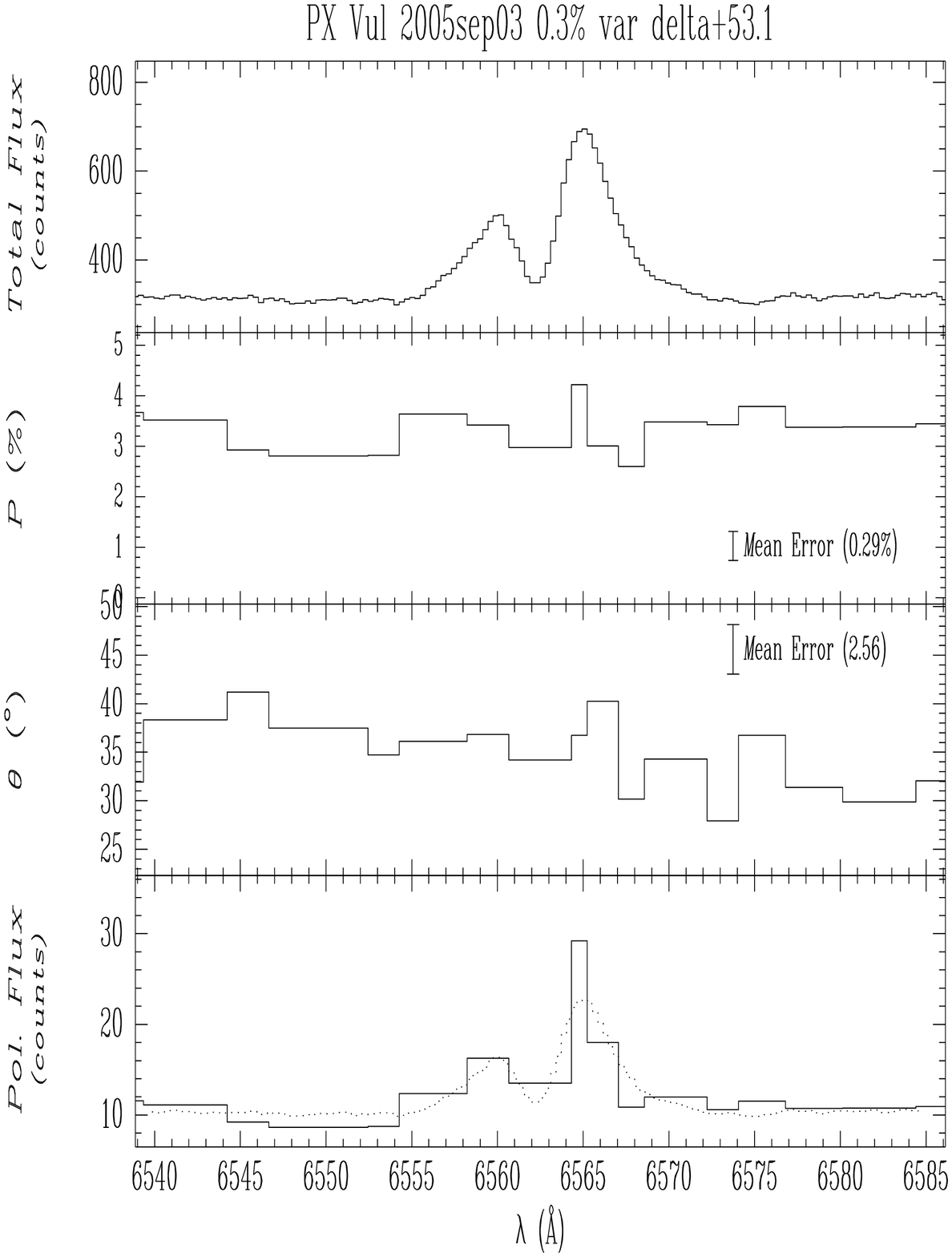}}
\linethickness{0.4pt}
\put (-142.5,210) {\line(0,-1){185}} 
\put (-83.5,210) {\line(0,-1){185}}
\linethickness{0.1pt} 
\put (-106.3,210) {\line(0,-1){185}}
}
\\
\resizebox{4.4cm}{!}{\includegraphics{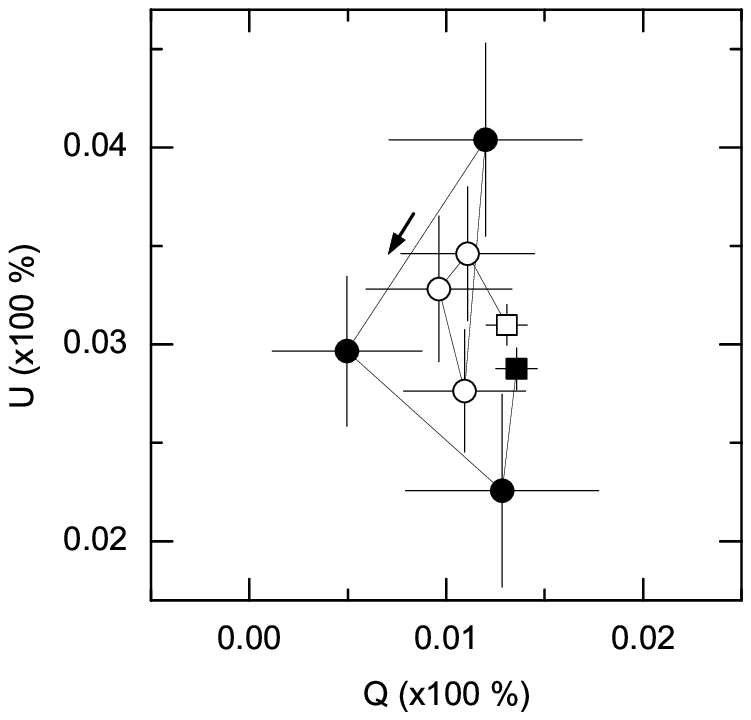}}
\\
\end{array}$
\caption{Spectropolarimetry of PV~Vul around H$\alpha$ on 2005/sep/03. \it{Top}\rm: The huge vertical lines indicate the full region (6555--6568~$\rm \AA$) where the line effect is evident. The thin vertical line indicates the limit (6563~$\rm \AA$) between the blue and red line effect. The spectra are as in Fig.~\ref{linerytau1} binned using a variable bin size with a constant polarization error (per bin) of 0.3$\%$. \it{Bottom}\rm: $\it{Q}$$-$$\it{U}$ diagram showing the line effect (6555--6563~$\rm \AA$ in white dots, 6563--6568~$\rm \AA$ in black dots). The direction in which the wavelength increases is also indicated (arrow) along with the mean values for the blue (white square) and red (black square) continuum. \textbf{The errors plotted around the line effect are assuming photon statistics directly.}} 
\label{linepxvul}
\end{figure}
\begin{figure}
\resizebox{8.8cm}{!}{\includegraphics{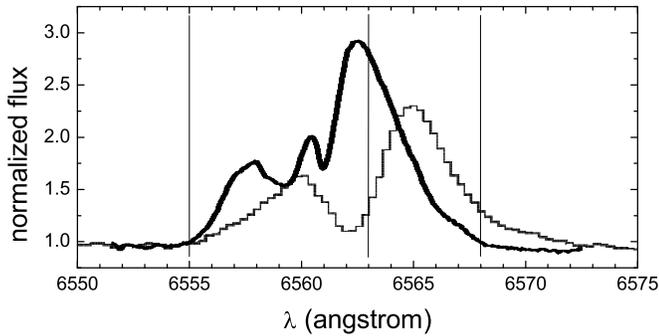}}
\caption{Comparison of the H$\alpha$ emission from PX Vul between 2004/jun (thick line, by Eisner et al. \cite{eis05} using Keck/HIRES) and 2005/sep (thin line, this work). The fluxes were normalized to the continuum in each case. The vertical lines are the same as in Fig.~\ref{linepxvul} to facilitate the comparison.}
\label{linecomppxvul}
\end{figure}

The spectropolarimetry of \object{PX~Vul} on 2005/sep is shown in Fig.~\ref{linepxvul}. The total flux of H$\alpha$ appeared as a double-peaked profile with the central dip at the continuum level. In relative fluxes, the red peak is $\sim$2.1 times more intense than the blue one. For comparison, H$\alpha$ shows a triple-peaked profile on 2004/jun (Eisner et al. \cite{eis05}) being the redder peak more intense than the other two (1.8 and 3.9 times, respectively). The bluer peak is not present in our data (1.3 year after) and spectral variability seems to be present on the H$\alpha$ emission changing from II$-$Bm to III$-$B type (Reipurth, Pedrosa \& Lago \cite{rei96}). This variability is also supported by the ratio between the redder peak and the continuum that fails from 2.9 (on 2004) to 2.3 (on 2005) (see Fig.~\ref{linecomppxvul}).

As in \S~\ref{rytau}, we detected the H$\alpha$ polarization line effect for \object{PX~Vul} using the $\it{Q}$$-$$\it{U}$ diagram. In this case, the main line effect is on the red peak (around three points on diagram between 6563--6568~$\rm \AA$) and appears as a loop consistent with the detection of a rotating disk. For a better comparison, we also plotted the points around the blue peak (6555--6563~$\rm \AA$) where another and subtle effect can be present. However, the strength of this effect is of the order of our errors and therefore this was not considered. 

The linear fit on the points that display the loop on the $\it{Q}$$-$$\it{U}$ diagram yields two possible solutions for the PA$_{\rm int}$ (91 and 91+90$\degr$, Table~\ref{tabresults}). As previous information is not available, the ambiguity on PA$_{\rm int}$ cannot be solved for \object{PX~Vul}. Following the sense of the PA rotation on the $\it{Q}$$-$$\it{U}$ diagram, the disk in \object{PX~Vul} rotates in a counter-clockwise direction as seen from the Earth. 


\section{Summary and Conclusions}
\label{concl}

We shown here the first results of an ongoing spectropolarimetric survey of YSOs gathered with EIFU$+$IAGPOL at LNA. All the line detections shown in this work are significant at least on one of the Stokes parameters with a 95\% confidence level.


The classical TTS \object{RY~Tau} was observed in 2004 and 2005. Interestingly, the double-peaked H$\alpha$ emission profile (II-B type) changed in $\sim$one year from a central dip at the continuum level to a dip just below the blue peak. \object{RY~Tau} faded about $\sim$0.8 mag between our observations with the blue/red peak ratio practically unchanged and without evidence of an anti-correlation between H$\alpha$ and the photometry as suggested by previous works.

Analysing the polarization spectra on the $\it{Q}$$-$$\it{U}$ diagram, \object{RY~Tau} shown a well-defined line effect passing from a loop on 2004 to a more complex pattern (loop$+$linear excursion) on 2005. Our observations are consistent with previous spectropolarimetric data from V03 where loop detection was reported. This feature is interpreted as a signature of a compact source for H$\alpha$$-$line photons. The intrinsic polarization PA obtained directly from the line effect did not seem change between the interval of our observation (PA$_{\rm int}$ $\sim$167$\degr$). We confirm that the disk in \object{RY~Tau} rotates clockwise. Considering the continuum polarization, a slight enhancement was observed on 2005 with the PA practically unchanged ($\sim$14$\degr$). The PA$_{\rm int}$ computed by the ISP corrected continuum method is more consistent with the presence of an optically thin disk in \object{RY~Tau}.  

For \object{PX~Vul}, the comparison with previous spectra shown that spectral variability is present on the H$\alpha$ emission passing from a II-Bm to II-B type in $\sim$1.3 yr. Nevertheless, the continuum polarization seems to be non-variable. We reported the detection of a rotating disk from the loop around H$\alpha$ line on the $\it{Q}$$-$$\it{U}$ diagram. The intrinsic polarization has a PA$\sim$91$\degr$ (with an ambiguity of 90$\degr$). An independent determination of the foreground polarization is necessary to better constrain the PA$_{\rm int}$. The sense of the rotation of the PA on the loop shows that the direction of the disk rotation of \object{PX~Vul} seems to be counter-clockwise. 

The finding of the signature of a loop in \object{RY~Tau} and \object{PX~Vul} confirms the trend of a compact source of line photons that is scattered off a rotating accretion disk in TTS. 

\begin{acknowledgements}
A. P. is thankful to FAPESP (grant 02/12880$-$0) and CNPq (DTI grant 382.585/07$-$03 associated to the PCI/MCT/ON program). A.M.M. acknowledges support from FAPESP and CNPq. Polarimetry at IAG-USP is supported by FAPESP grant 01/12589$-$1. 

\end{acknowledgements}


\end{document}